\documentclass[a4paper]{article}

\usepackage[utf8]{inputenc} 

\usepackage{amsfonts}
\usepackage{authblk}
\usepackage{amssymb}
\usepackage{amsbsy}
\usepackage[margin=2cm]{geometry}

\usepackage{xcolor}
\usepackage{cite}
\usepackage{mathrsfs}
\usepackage{hyperref}
\usepackage{graphicx} 

\usepackage{booktabs} 
\usepackage{array} 
\usepackage{paralist} 
\usepackage{verbatim} 

\usepackage{subfigure}
\makeatletter
\renewcommand{\@thesubfigure}{\normalsize(\textbf{\alph{subfigure}})}
\makeatother

\usepackage{fancyhdr} 
\title{Geometry and Entanglement of Two-Qubit States
in the Quantum Probabilistic Representation}

\author[1,2]{Julio A. L\'opez-Sald\'ivar}
\author[1]{Octavio Casta\~nos}
\author[1]{Eduardo Nahmad-Achar}
\author[1]{Ram\'on~L\'opez-Pe\~na}
\author[3]{Margarita A. Man'ko}
\author[2,3,4]{Vladimir I. Man'ko}
\affil[1]{Instituto de Ciencias Nucleares, Universidad Nacional Aut\'onoma de M\'exico,
Apdo. Postal 70-543, Ciudad de México 04510, M\'exico}
\affil[2]{Moscow Institute of Physics and Technology (State University), Institutskii per. 9,
Dolgoprudnyi, Moscow~Region 141700, Russia}
\affil[3]{Lebedev Physical Institute, Russian Academy of Sciences, Leninskii Prospect 53,
Moscow 119991, Russia}
\affil[4]{Department of Physics, Tomsk State University, Lenin Avenue 36, Tomsk 634050, Russia}
\date{}

\begin{document}
\maketitle
\begin{abstract}
A new geometric representation of qubit and qutrit states based on
probability simplexes is used to describe the separability and entanglement
properties of density matrices of two qubits. The Peres--Horodecki  positive partial transpose (ppt)-criterion and the concurrence inequalities are formulated as the conditions that the introduced
probability distributions must satisfy to present entanglement. A four-level system, where one or two states are inaccessible, is considered as an example of applying the elaborated probability approach in an explicit form. The areas of three
Triadas of Malevich's squares for entangled states of two qubits are defined
through the qutrit state, and the critical values of the sum of their areas
are calculated. We always find an interval for the sum of the square
areas, which provides the possibility for an experimental checkup of the
entanglement of the system in terms of the probabilities. 
\end{abstract}


\section{Introduction}
The states of quantum systems are determined by wave
functions~\cite{Schroed1926,Schroed19261} (pure states) or density
matrices~\cite{Landau27,vonNeumann27}. The corresponding definition of these
states is done by using state vectors or density operators in the Hilbert
space~\cite{Dirac-book}. For qudits, we discuss the
approach where the quantum states are identified with fair probability
distributions. Different quasiprobability representations of the density
operators, such as the Wigner function~\cite{wigner32}, Husimi
$Q$-function~\cite{husimi} or the Glauber--Sudarshan
$P$-function~\cite{glauber63,sud63}, were introduced to describe continuous variable quantum systems.
These functions have been also defined for discrete variable systems such as spin-1/2
particles~\cite{stratonovich57}. In addition, the formulation of quantum states without
probability amplitudes was proposed in~\cite{wooters2}, and the geometric
definition of the quantum state determined by the transition probabilities was
presented in~\cite{mielnik}.

Recently, the probability representation of quantum states was introduced both
for continuous variables~\cite{tombesi96} and spin
systems~\cite{dodonov1,OlgaJETP}. This approach uses quantum tomograms, which
can be measured in experiments as the prime objects identified with the
quantum state of an arbitrary system. The qubit or spin-1/2 state, within the
framework of the tomographic probability representation, is~identified with
the set of  three probability distributions of spin projections on three
perpendicular directions in the space. This description of the qubit state was
studied and illustrated by the triangle geometry of the system, using
the so-called Malevich square representation~\cite{chernega17a} known also as quantum suprematism
approach (after the Russian painter Kazmir
Malevich (1879--1935), founder~of suprematism, an art movement started around 1913 focused on basic geometric figures). Such a geometric representation provides the picture of the qubit
state in terms of three squares on the plane obtained through an invertible
map of the points in the Bloch sphere onto the probability distributions. This
approach has been extended for qutrit
states~\cite{chernega17b,chernega17c,chern2} and, in~principle, was
generalized to qudit states. An important role of symmetries and group
representations, in particular, for spin states was reviewed
in~\cite{marmo-book}. Within the framework of the geometric formulation of
quantum mechanics~\cite{ZyczkowskiBengson,marmo-book}, an explicit
construction of the Fisher--Rao tomographic metric for qubit and qutrit
density matrices is established in a quorum of reference
frames~\cite{marmo,marmo1}. In addition, using the same approach, the volume of two-qubit states which have maximal random subsystems (where the reduced density matrices $\hat{\rho}_{1,2}=\hat{I}/2$), has been calculated in~\cite{mancini} as a function of the purity of the composite system.

Quantum computers manipulate qubits by operations based on Pauli matrices; we
elaborate in this work  the decomposition of qutrit states into qubit states
and hope that the proposed decomposition will also allow the manipulation of
qutrits and, in general, of qudits in quantum computing algorithms. An example
of the mapping of oscillator creation and annihilation operators onto qubits
using the Jordan--Schwinger map~\cite{Jordan,Schwinger} and manipulation of
the qubits in context of information technologies has recently been given in
\cite{dumitrescu18}.

The aim of this paper is to study, within the probability representation of
quantum
states~\cite{tombesi96,dodonov1,OlgaJETP,chernega17a,chernega17b,chernega17c,chern2,Filip-JRLR,Ocasta-jpa2004,MA-entropy} reviewed in~\cite{MarmoPS150},
the triangle geometry, separability, and entanglement of a composite system of
two qubits in specific states. In addition, we elaborate the description of the state
quantumness by finding new bounds for qubit and qutrit state characteristics
presented in terms of square areas given by probability distributions associated with the triangle geometry of their
states. It is worth noting that the classical probability distributions and
their interference were discussed within the framework of the state vectors in
Hilbert space by Khrennikov~\cite{KhrPr,KhrFPh,KhrQI,khrebook}. Here, the interference is a feature of multi-contextuality. This is not only a problem of classical versus quantum probability, but also quantum versus general contextual probability. The superposition
principle for spin-1/2 state vectors was presented in explicit form as the
nonlinear superposition of the classical probability distributions determining
the qubit states in~\cite{Chern1,MAPS,chern2}. This superposition was
illustrated geometrically in the quantum suprematism approach as a superposition
of squares. The approach called the suprematism in art is described
in~\cite{Shatskikh}. It is worth noting that a methodological relation
of the Malevich black square with effectiveness for experimental tools in
physics was mentioned in~\cite{zeilinger}.

The system of two qubits can be realized as a system of two two-level atoms;
this system has four levels. Specific states of the four-level system are the
states where either one level or two levels of the four are not occupied. It
means that some states from the set of possible states are inaccessible. We~
discuss the properties of such states for two-qubit systems. Thus, we study,
within the probability representation, the triangle geometry and separability
of the specific states of two qubits. This is done by considering that one or
two of the composite two-qubit states are not available, which yields
to concurrences depending only on two probability distributions of dichotomous
random variables. Note~that, when there is only one inaccessible state, an
additional nonlinear mapping suggested in~\cite{chernega17c} needs to be
applied to determine the geometric picture of the states in terms of three
triads of squares. The Peres--Horodecki criterion
\cite{peres,horodecki} is used to establish the separability or entanglement
properties of the two-qubit states.

We point out the following aspects of our approach. The entanglement in a
two-qubit system is completely a quantum phenomenon. In view of this fact, it
seems to be necessary to use for its description mandatory ingredients such as
Hilbert spaces, vectors in the Hilbert space, and density operators acting in
the space. As we  demonstrate, and it is our goal, it is possible to
describe this quantum phenomenon making the identification of qubit states
with fair classical-like measurable probabilities. Our conjecture is that
other completely quantum phenomena in some other systems such as quantum
correlations (e.g., Bell correlations) can also be formally described using
the states identification with probability distributions.

This paper is organized as follows.

In Section~\ref{S2}, a short review of the qubit and qutrit state probabilistic
description given in the quantum suprematism geometric representation is
presented. In Section~\ref{S3}, two-qubit states both separable and entangled are
considered in the probability representation when there are one or two
inaccessible states. Section~\ref{S4} presents an example in which the inequalities over the square areas and over the sum of areas lead to conditions which can be used for controlling measurement processes. Conclusions and perspectives are presented in Section~\ref{S5}.

\section{Qubit and Qutrit States in Quantum Geometric Representation}
\label{S2}
In this section, we review how the Bloch sphere geometry of qubit states is
mapped onto a triangle geometry of qubit and qutrit states. The construction of
the map is described in terms of the measurements of probabilities along the
quorum of reference frames ~\cite{chernega17a,chernega17b,chernega17c}.

\subsection{Qubit Case}

We start with a qubit density matrix $\hat{\rho}=\hat{\rho}^\dagger$, ${\rm Tr}(\hat{\rho})=1$ satisfying the nonnegativity condition of its eigenvalues,
i.e.,
\begin{equation}
\hat{\rho}=\left( \begin{array}{cc}
\rho_{11} & \rho_{12} \\
\rho_{21} & \rho_{22}
\end{array} \right) \, , \quad \rho_{21}=\rho_{12}^* \, , \quad \rho_{11}+\rho_{22}=1 \, ,
\label{eq1}
\end{equation}
and
\begin{equation}
\rho_{11}\, \rho_{22}-\rho_{12} \,\rho_{21} \geq 0 \, .
\end{equation}

The matrix elements $\rho_{jk}$; $j,k=1,2$ may be constructed in terms of
three probability distributions $\boldsymbol{\mathcal{P}}_1=(p_1,1-p_1)$, $\boldsymbol{\mathcal{P}}_2=(p_2,1-p_2)$, and $\boldsymbol{\mathcal{P}}_3=(p_3,1-p_3)$, where $0 \leq p_k \leq 1$; $k=1,2,3$
are probabilities of spin-1/2 projections $m= \pm 1/2$ along the axes $x, y, z$,
respectively. Each probability is related to the expectation values of the
projectors
\begin{equation}
\hat{\rho}_1=\frac{1}{2}\left( \begin{array}{cc} 1 & 1 \\ 1 & 1 \end{array}
\right),\quad \hat{\rho}_2=\frac{1}{2}\left( \begin{array}{cc} 1 & -i \\ i & 1
\end{array} \right),\quad  \hat{\rho}_3=\left( \begin{array}{cc} 1 & 0 \\ 0 & 0
\end{array} \right) \, ,
\end{equation}
defining the probabilities ${\rm Tr}(\hat{\rho}\hat{\rho}_k)=p_k$ which can be
measured experimentally. These measurements allow  reconstructing~Equation (\ref{eq1})
in the form
\begin{equation}
\hat{\rho}=\left( \begin{array}{cc}
p_3 & p_1-1/2-i(p_2-1/2) \\
p_1-1/2+i(p_2-1/2) & 1-p_3
\end{array}\right) .
\label{rqubit}
\end{equation}

Note that $p_1$, $p_2$, and $p_3$ are classical probabilities of measuring the
projection of angular momentum $m=1/2$ in three different reference frames. We
point out that, for a system of three independent classical coins, its
statistics are also described by the same three probabilities.

\begin{figure}[!]
\centering \subfigure[] {\includegraphics[scale=0.25]{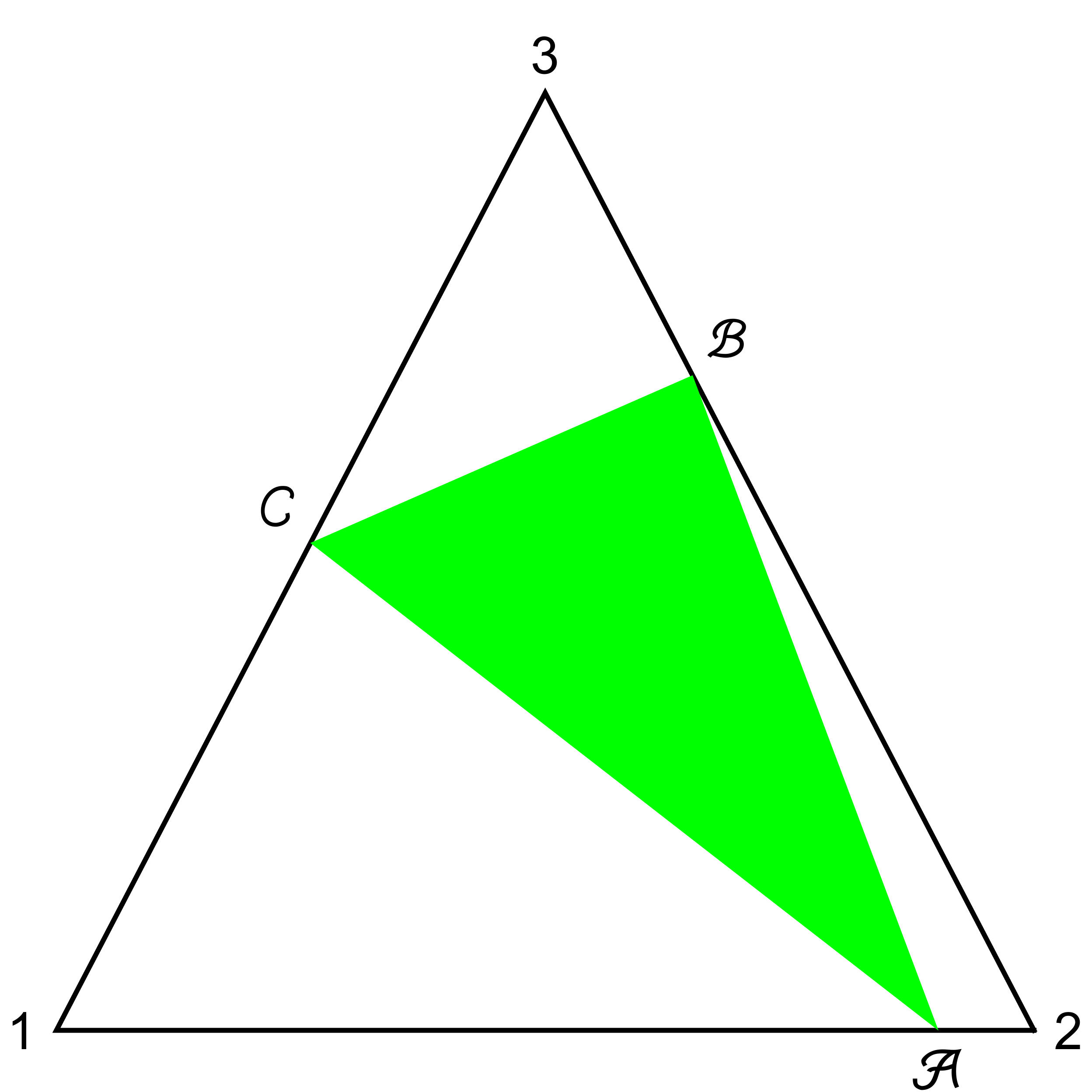}}
 \subfigure[] {\includegraphics[scale=0.25]{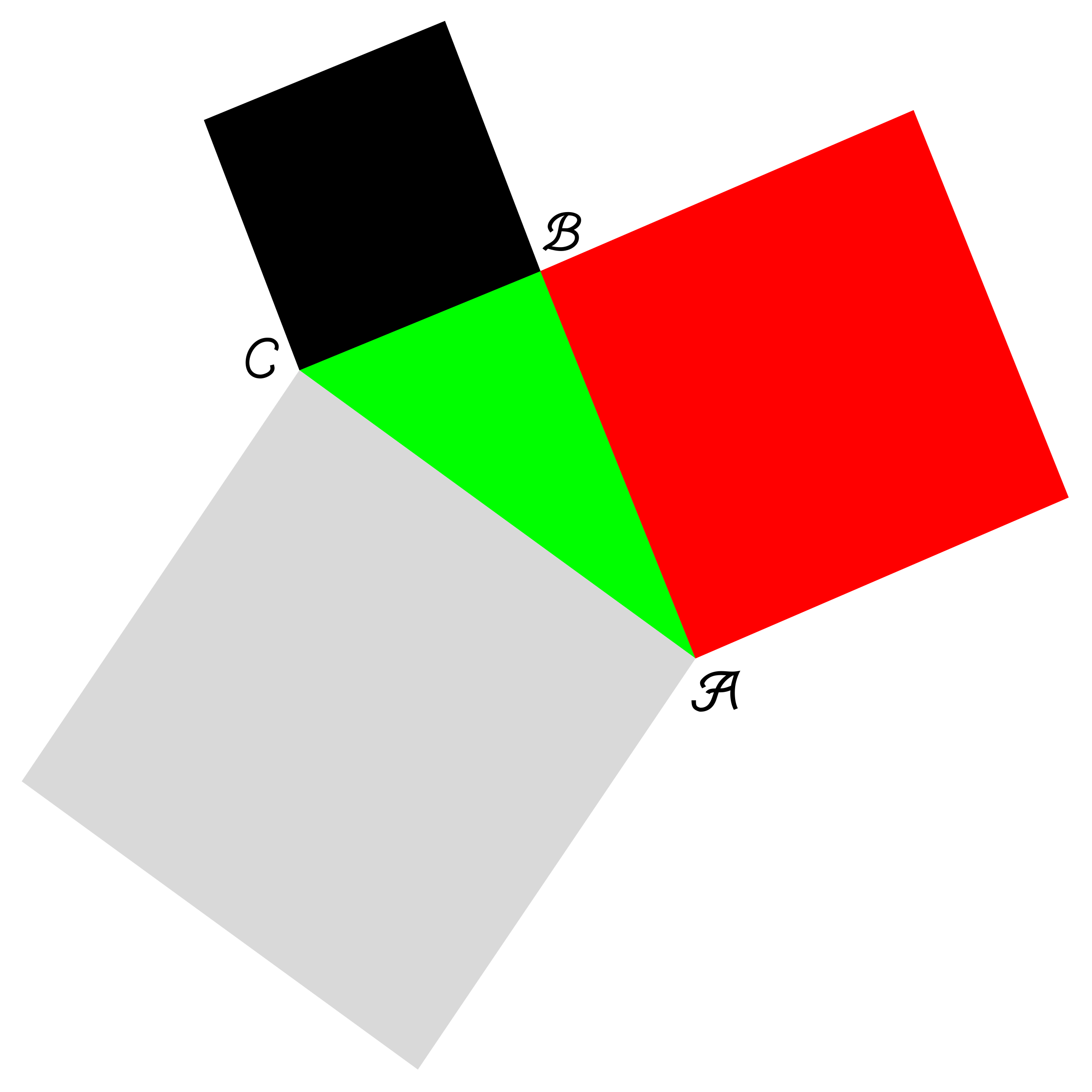}} \vspace{-8pt}
\caption{(\textbf{a}) Triangle representation of
the qubit state by three points along the perimeter of an equilateral triangle
of side length $\sqrt{2}$; and (\textbf{b}) Malevich's squares associated to the state.}
\label{triangle}
\end{figure}

A state with the density matrix $\hat{\rho}_k$, as described above, has spin
projections $m=\pm 1/2$ on the three perpendicular directions $x, y, z$. This means that the state $\hat{\rho}$ is identified with three
probabilities $p_1$, $p_2$, and $p_3$. The nonnegativity of the density matrix
$\hat{\rho} \geq 0$ provides the condition
\begin{equation}
(p_1-1/2)^2+(p_2-1/2)^2+(p_3-1/2)^2\leq 1/4 \, ,
\label{cons}
\end{equation}
i.e., there exist quantum correlations between the spin projections on the
perpendicular directions $x, y, z$. In contrast, for three classical coins
described as the probability vectors $\boldsymbol{\mathcal{P}}_1$,
$\boldsymbol{\mathcal{P}}_2$, and $\boldsymbol{\mathcal{P}}_3$, there are no
constraints~(Equation \ref{cons}). The endpoints of the probability vectors
$\boldsymbol{\mathcal{P}}_k$ with components $p_k$ and $1-p_k$ are situated
along $1$-simplexes, which form the hypotenuse of rectangular triangles of
side $1$. If one connects the hypotenuses, one can obtain an equilateral
triangle with side length $\sqrt{2}$ (see~\cite{chernega17a}). Then, the state of a qubit can be
represented by three points along the triangle sides, as shown in
Figure~\ref{triangle}; some new entropic inequalities were obtained for qubit
systems in \cite{JulioPhysicaA}.

$\mathcal{A}$, $\mathcal{B}$, and $\mathcal{C}$ show the endpoints of vectors
$\boldsymbol{\mathcal{P}}_1$, $\boldsymbol{\mathcal{P}}_2$, and
$\boldsymbol{\mathcal{P}}_3$ on the simplexes. The side lengths  $l_k$, $k=1,2,3$ of the triangle
$\triangle(\mathcal{A} \mathcal{B} \mathcal{C})$ can be expressed in terms of
probabilities as follows:
\begin{equation}
l_k=(2 p_k^2+2 p_{k+1}^2+2 p_k p_{k+1}-4 p_k-2p_{k+1}+2)^{1/2} \, .
\end{equation}

From these, one can define three squares with sides $l_1$, $l_2$, and $l_3$. The triad of squares illustrates the qubit density matrix, and it has a one-to-one
correspondence with the Bloch parameters of the state
\begin{equation}
x=2\, p_1-1 \, , \quad y=2\, p_2 - 1 \, , \quad z=2\,p_3 -1 \, .
\end{equation}

The linear relation between the probabilities and the Bloch vector parameters,
together with the condition~in~Equation (\ref{cons}), allows an analogous construction to
the Bloch sphere with center at $p_k=1/2$; $k=1,2,3$ and radius $1/2$. In this
representation, the most mixed state with density operator
$\hat{\rho}=\mathbf{I}/2$ is located at the center of the sphere, and one can
find the pure states on the surface.

The sum of the square areas is given in terms of the triangle
lengths as $S=l_1^2+l_2^2+l_3^2$ and explicitly in terms of the probabilities
as
\begin{equation}
 S(p_1, p_2, p_3)= 2 \left(2 {p_1}^2+3 (1-{p_1}-{p_2}-{p_3})+{p_1}
 {p_2}+{p_1} {p_3}+2 {p_2}^2+ {p_2} {p_3}+2 {p_3}^2\right).
\label{area1}
\end{equation}

The difference with the classical treatment with three coins is that the
uncertainty relation~in~Equation (\ref{cons}) is not imposed. In this classical case, the
sum of the square areas satisfies the inequality
\begin{equation}
3/2 \leq S_{c}\leq 6\ ,
\end{equation}
where the lower bound corresponds to the probabilities $p_{1}=p_{2}=p_{3}=1/2$
and the upper limit, to $p_{1}=p_{2}=p_{3}=1$.

For the quantum case of the qubit state, one has to consider the
constraint~in~Equation (\ref{cons}). For pure states, i.e.,~when the equality is satisfied
in the uncertainty relation~in~Equation (\ref{cons}), the sum of the square areas takes local maxima with value $S_{q}=9/4$  and two global maxima with
$S_{q}=3$. The lower bound, $S_{q}=3/2$, is given by the maximum mixed states.
Therefore, in the quantum case, the sum of the square areas~satisfies
\begin{equation}
3/2 \leq S_{q}\leq 3\ .
\label{mal_qubits}
\end{equation}

For $S_{q}=3$, the triangle $\mathcal{A}\mathcal{B}\mathcal{C}$ is equilateral with the side length equal to 1, and for $S_{q}=3/2$, the~equilateral
triangle $\mathcal{A}\mathcal{B}\mathcal{C}$ has the side length equal to
$\sqrt 2/2$.

\begin{figure}[!]
\centering
\includegraphics[scale=0.35]{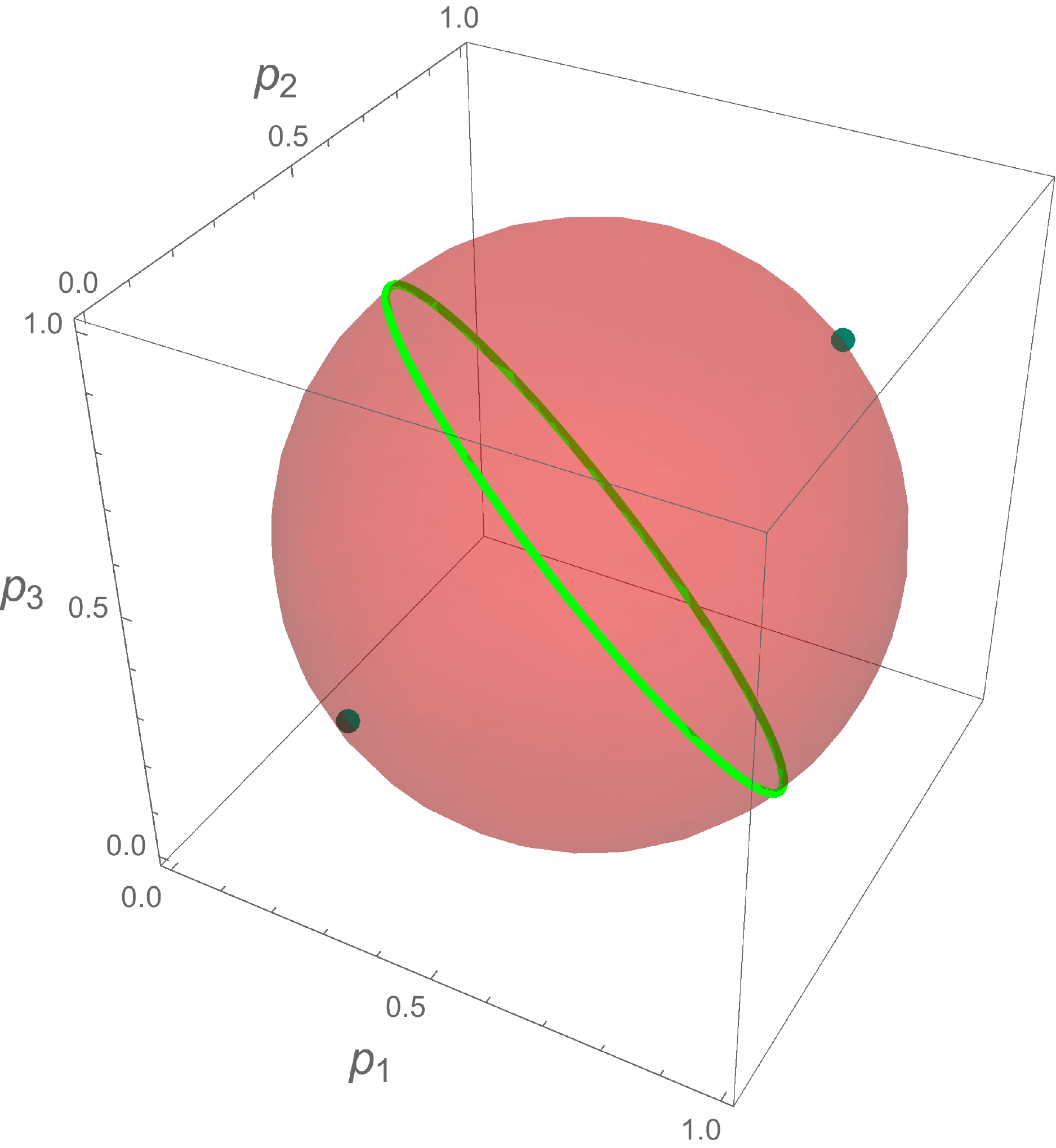}
\caption{Geometric interpretation of a qubit in the probability
representation. The (red) sphere is centered at the maximum mixed state and
has radius $1/2$. The~great circle is associated to pure states, where
$S_{q}=9/4$, and the dots are pure states, where $S_{q}=3$.}
\label{pure_states1}
\end{figure}

In Figure~\ref{pure_states1}, we show the geometric interpretation of the qubit
state in the probability representation, together with the pure states that
maximize the sum of the triad areas $S_{q}$ in the quantum
case. The great circle determined by the points
\begin{equation}
p_{2}=\frac{1}{4} \left(3 - 2 p_{1} + \sqrt{-1 + 12\, p_{1} - 12\,
p_{1}^{2}}\right),\quad p_{3}=\frac{1}{4} \left(3 - 2 p_{1} - \sqrt{-1 + 12
p_{1} - 12 p_{1}^{2}}\right),
\end{equation}
where $(3-\sqrt{6})/6\leq p_{1}\leq(3+\sqrt{6})/6$, corresponds to local maxima.
The absolute maxima are reached at the probability vectors
\begin{equation}
\Big(p_{1},\,p_{2},\,p_{3}\Big)=\Big(\frac{1}{6}(3-\sqrt{3}),\,
\frac{1}{6}(3-\sqrt{3}),\, \frac{1}{6}(3-\sqrt{3})\Big),\ \
\Big(\frac{1}{6}(3+\sqrt{3}),\quad \frac{1}{6}(3+\sqrt{3}),\,
\frac{1}{6}(3+\sqrt{3})\Big)\ .
\end{equation}

In addition to the areas, the linear entropy of the system can be calculated using the relation
\begin{equation}
{\cal S}_L=2\sum_{j=1}^3 p_j (1-p_j) -1\, .
\label{sl}
\end{equation}

It is important to note that, if $p_j$ represents the standard probability
distribution corresponding to a dichotomous random variable (e.g., a coin),
the terms $\eta_j=p_j (1-p_j)$ measure the fairness of the system. If the
dichotomous variable has the same probability for both categories
$p_j=1-p_j=1/2$, then $\eta_j=1/4$ constitutes the maximum fairness situation.
In the opposite case, when one of the categories of the dichotomous variable
is not possible, the fairness has a minimum $\eta_j=0$.

One can see that, for maximum fairness, the qubit state corresponds to the
most mixed state $\hat{\rho}=\mathbf{I}/2$ and has a linear entropy ${\cal
S}_L=1/2$. When one has minimum fairness, there exist two possibilities:
$p_j=0$ and $p_j=1$. At any of those values, the linear entropy has a value of
$-1$ which is not physical, so one can conclude that the probabilities $p_j$
cannot be zero at the same time, nor can they all be equal to $1$ or any
combination of $0$ and $1$, in order to represent the qubit state. As can be
seen in Figure~\ref{pure_states1}, those points are located outside the
permitted sphere given by~Equation (\ref{cons}).

In addition, the linear entropy of the system is proportional to the sum of the
squared lengths of the triangles $T_1=\triangle{(\mathcal{A} \mathcal{B} 2)}$,
$T_2=\triangle{(\mathcal{B} \mathcal{C}3 )}$, and
$T_3=\triangle{(\mathcal{A}\mathcal{C} 1)}$, i.e., $\sum_{j=1}^3
2((1-p_j)^2+p_{j+1}^2)+l_j^2$, minus the squared lengths of the triangle
$T_4=\triangle{(\mathcal{A}\mathcal{B}\mathcal{C})}$, i.e., $\sum_{j=1}^3
l_j^2$; explicitly,
\begin{equation}
{\cal S}_L=2-\sum_{j=1}^3 [(1-p_j)^2+p_{j+1}^2] \, ,
\label{sl1}
\end{equation}
where $p_4=p_1$. Note that ~Equations~(\ref{sl}) and (\ref{sl1}) are
equivalent.

\subsection{Qutrit Case}
The probabilistic representation of the qubit state can also be extended to
higher dimensions. We~consider the example of  the qutrit state.  The density
matrix of the qutrit state
\begin{equation}
\hat{\rho}_3 = \left( \begin{array}{ccc}
\rho_{11} & \rho_{12} & \rho_{13} \\
\rho_{21} & \rho_{22} & \rho_{23} \\
\rho_{31} & \rho_{32} & \rho_{33}
\end{array}\right)\, ,
\end{equation}
can be described using the eight generators of the su(3) algebra represented
by the Gell--Mann matrices~\cite{gellmann} $\hat{\lambda}_1, \ldots,
\hat{\lambda}_8$, i.e.,
\[
\hat{\rho}_3=\frac{1}{3}\,\hat{I}+\frac{1}{2}\, \sum_{j=1}^8 a_j
\hat{\lambda}_j \, ,
\]
where $a_j\in\mathbb{R}$ are the entries of the generalized Bloch vector.
Amongst the Gell--Mann matrices, there~exist three sets of operators which form
su(2) algebras, viz., $\{\hat{\lambda}_1, \hat{\lambda}_2, \hat{\lambda}_3
\}$, $\{\hat{\lambda}_4, \hat{\lambda}_5,
(\hat{\lambda}_3+\sqrt{3}\,\hat{\lambda}_8)/2\}$, and~$\{\hat{\lambda}_6,
\hat{\lambda}_7, (-\hat{\lambda}_3+\sqrt{3} \, \hat{\lambda}_8)/2 \}$. Given
this property, one can think of a possible definition of qubit states using
these three sets of operators. An algorithmic procedure to define qubit states
is the following: The matrix $\hat{\rho}_3$ is first extended to two
4$\times$4 density matrices, where one row and one column are equal to zero,
as follows:
\[
\hat{\rho}_1=\left(\begin{array}{cc}
\hat{\rho}_3 & 0 \\
0 & 0
\end{array} \right) \, , \quad  \hat{\rho}_2=\left(\begin{array}{cc}
0 & 0 \\
0 & \hat{\rho}_3
\end{array} \right) \, .
\]

\begin{figure}[!]
\centering
\includegraphics[scale=0.80]{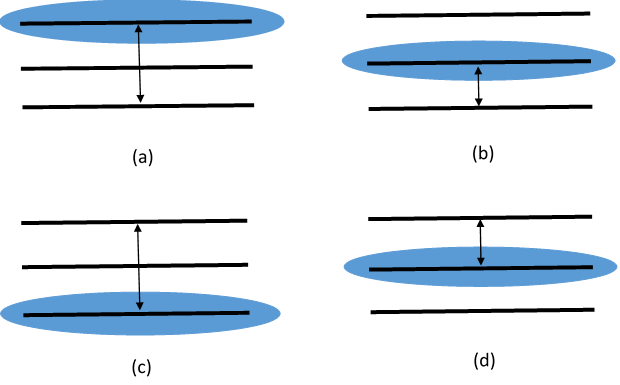} \vspace{-6pt}
\caption{Schematic representation of qubits defined by a generic
three-level system given by the density
matrices: (\textbf{a}) $\hat{\rho}^{(A)}$; (\textbf{b}) $\hat{\rho}^{(B)}$; (\textbf{c})
$\hat{\rho}^{(C)}$; and (\textbf{d}) $\hat{\rho}^{(D)}$. In all cases, the occupation number of the states in blue
define the diagonal terms, while the arrows denote the transitions which
define the off-diagonal terms of the qubits.}
 \label{qbits}
\end{figure}

 Interpreting the resulting matrices as density operators for two
qubit systems, we make use of the partial trace operation to define four
matrices that must be positive semidefinite $\hat{\rho}^{(A)}$,
$\hat{\rho}^{(B)}$, $\hat{\rho}^{(C)}$, and $\hat{\rho}^{(D)}$, which are not
independent
\begin{eqnarray}
 \hat{\rho}^{(A)}=\left(\begin{array}{cc}
 1-\rho_{33} & \rho_{13}  \\
 \rho_{31} & \rho_{33}
 \end{array}\right) \, , \qquad
  \hat{\rho}^{(B)}=\left(\begin{array}{cc}
 1-\rho_{22} & \rho_{12} \\
 \rho_{21} & \rho_{22}
 \end{array}\right) \, , \nonumber \\
 \hat{\rho}^{(C)}=\left(\begin{array}{cc}
 \rho_{11} & \rho_{13} \\
 \rho_{31} & 1-\rho_{11}
 \end{array}\right) \, , \qquad
  \hat{\rho}^{(D)}=\left(\begin{array}{cc}
 \rho_{22} & \rho_{23} \\
 \rho_{32} & 1-\rho_{22}
 \end{array}\right) \, ; \label{qubitss}
 \end{eqnarray}
 \noindent in Figure~\ref{qbits}, it is shown that associated to any of the qubit density matrices in Equation~(\ref{qubitss}) is a three-level system. In each case, the population of one of the levels with the transition probability to another level determine different qubits. It can be seen that the
off-diagonal components of the matrices in~Equation  (\ref{qubitss}) are naturally arranged in the sets
given by the su(2) algebras, i.e., $A:\{a_4, a_5\}$, $B:\{a_1, a_2 \}$,
$C:\{a_4, a_5\}$, and $D:\{a_6, a_7\}$. Therefore, each one of these density
matrices can be decomposed in terms of three probabilities as described in
Equation~(\ref{rqubit}). Choosing the independent qubits as $\hat{\rho}^{(A)}$,
$\hat{\rho}^{(B)}$, $\hat{\rho}^{(D)}$, one can retrieve the original 3 $\times
$ 3 density matrix in the form
\begin{equation}
\hat{\rho}_3 = \left( \begin{array}{ccc}
p_3^{(A)}+p_3^{(B)}-1 & B & A \\
B^* & 1-p_3^{(B)} & D \\
A^* & D^* & 1-p_3^{(A)}
\end{array}\right) \, ,
\label{qutrit}
\end{equation}
where $A=p_1^{(A)}-1/2-i(p_2^{(A)}-1/2)$, $B=p_1^{(B)}-1/2-i (p_2^{(B)}-1/2)$,
and $D=p_1^{(D)}-1/2-i(p_2^{(D)}-1/2)$; here, the numbers
$p_{1,2,3}^{(A),(B),(D)}$ are probabilities satisfying the
inequality~in~Equation (\ref{cons}). It is worth mentioning that qubits
$\hat{\rho}^{(B)}$, $\hat{\rho}^{(C)}$, and $\hat{\rho}^{(D)}$ can also be
used to describe the system, as shown below.

The qubit probabilities can be obtained in terms of the tomographic
probabilities used for the state reconstruction. It is known that, to
reconstruct the qutrit state, one needs to measure the probabilities
corresponding to the spin projections $m=0,1$ on the $z$ axis in four
different reference frames. Each of these frames constitute a general rotation of the
density matrix acting by the operator $\hat{U}=\prod_{j=1}^8 \exp (i \theta_j
\hat{\lambda}_j)$ on the original state $\hat{U}^\dagger \hat{\rho}_3
\hat{U}$.

As in the qubit case, the linear entropy of the system can be obtained
as
\begin{equation}
{\cal S}_L=2\left( \sum_{j=A,B,D}\sum_{k=1}^3 p_k^{(j)}\left(1-p_k^{(j)}\right)
+ p_3^{(A)}\left(1-p_3^{(B)}\right)+ p_3^{(B)2}\right)-5 \, ,
\end{equation}
\noindent with $p_3^{(D)}=1-p_3^{(B)}$. Even though the
expression is similar, one can see that, in addition to the fairness terms for
each probability $\eta_{jk}=p_k^{(j)}(1-p_k^{(j)})$, we also have the joint
probability distribution $p_3^{(A)}(1-p_3^{(B)})$, and the probability
$p_3^{(B)}$. It can be shown that ${\cal S}_L^{(A)}+{\cal S}_L^{(B)}+{\cal
S}_L^{(D)}=2\sum_{j=A,B,C}\sum_{k=1}^3 p_k^{(j)}\left(1-p_k^{(j)}\right)-3$,
so the linear entropy is expressed as
\begin{equation}
{\cal S}_L=\sum_{j=A,B,D} {\cal S}_L^{(j)}-2 \left(1-p_3^{(B)}\right) \left(1+p_3^{(B)}-p_3^{(A)}\right) \, ,
\label{purity}
\end{equation}
\noindent which can be obtained geometrically, in view of the property of the
entropy for the three qubits $A$, $B$, and $D$ in terms of the squared lengths
of the triad squares, as discussed previously.  It is important to
note that, in general, the sum of the linear entropies for qubits is larger
than the linear entropy of the qutrit, i.e., ${\cal S}_L \leq \sum_{j}{\cal
S}_L^{(j)}$.

Given the nonnegativity of the qutrit density matrix, there exist correlations
between its matrix components, i.e., if a change in the system is done, these
components must change in a way to guaranty the hermiticity and nonnegativity of
the state. Even if we might be able to change a single matrix element of the
state, a change in all the others would take place after. These correlations
also imply a correlation between the component qubits defined above. For these
reasons, one can think of Equation~(\ref{purity}) as a way to measure correlations
between different components of the qutrit state, that is between different qubits.

Next, we determine the bounds associated to the sum of the square areas
for the qutrit in the $B,C,D$ qubit representation,
\begin{equation}
S= S\left(p^{(B)}_1, \, p^{(B)}_2, \, p^{(B)}_3\right) + S\left(p^{(C)}_1,\,
p^{(C)}_2, \, p^{(C)}_3\right) + S\left(p^{(D)}_1, \, p^{(D)}_2, \, 1 -
p^{(B)}_3\right) \, .
\label{sums}
\end{equation}

\begin{figure}[!]
\centering
\includegraphics[scale=0.35]{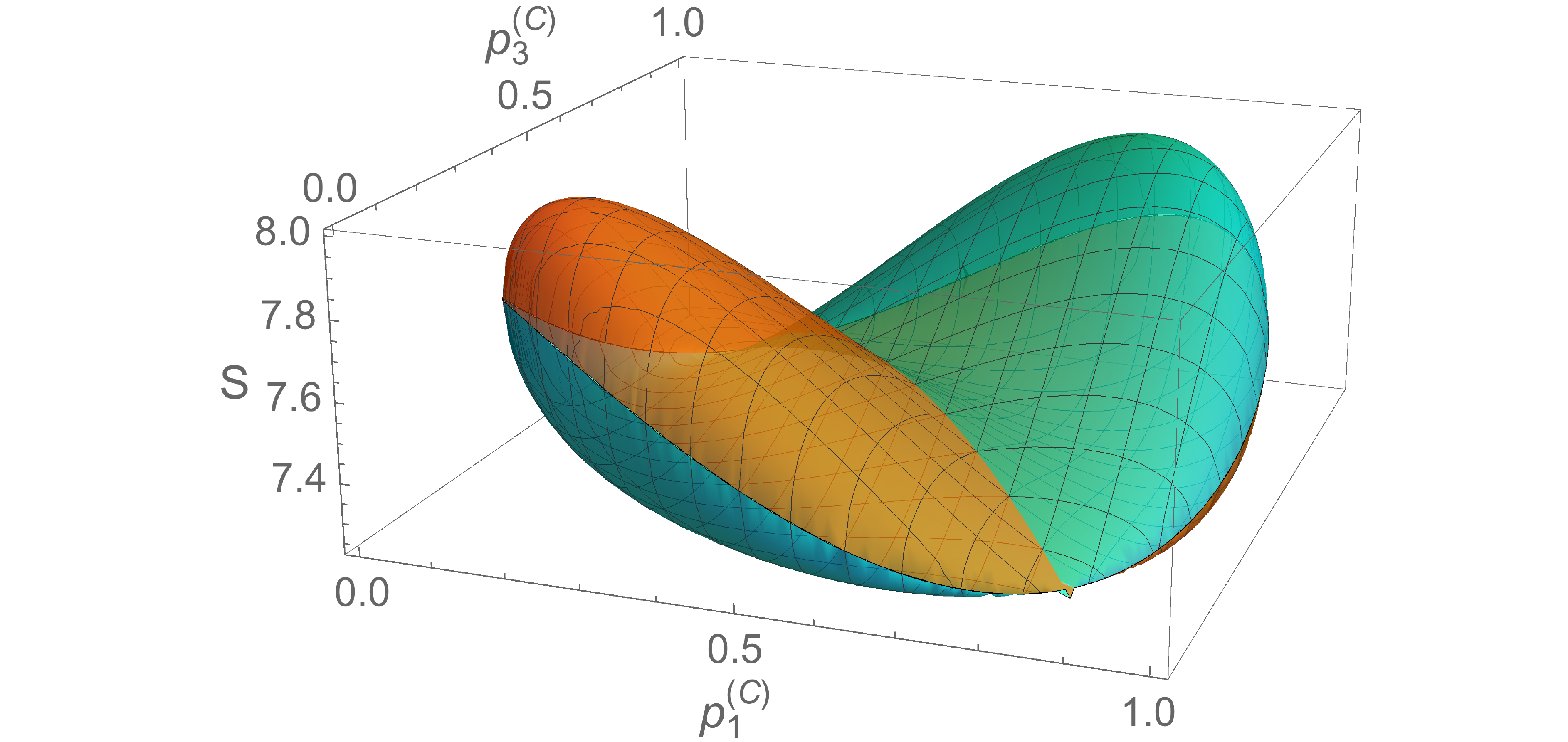}
\caption{The pure qubit representation of the sum of the square areas
in the probability space of $(p^{(C)}_1,p^{(C)}_3)$. It corresponds to pure
qutrit states. Each color denotes independent solutions.} \label{area2}
\end{figure}

We demonstrated that the qutrit density matrix can be written in terms of
eight probabilities establishing a three-qubit representation.  By requiring
the purity of the qutrit and the fact that qubits correspond also to pure
states, one can reduce the number of free probabilities to $p^{(C)}_1$ and
$p^{(C)}_3$. The~minimum value of the sum of the square areas is
obtained when all the probabilities take the value $1/2$, which corresponds to
a diagonal density matrix for the qutrit, diag$(1/2,1/2,0)$.  The maximum
value for the qutrit in the pure qubit representation reads $S=8$, while the
minimum is $29/4$. The~region of $(p^{(C)}_1,p^{(C)}_3)$ formed with pure
qubit states is the surface shown in Figure~\ref{area2}. The extreme bounds are given by
\begin{equation}
\frac{9}{2} \leq S \lesssim 8.1565 \, ,
\label{bb}
\end{equation}
where the upper bound is associated with the pure state and the probabilities $p^{(B)}_1\approx
0.5733$, $p^{(B)}_2\approx 0.5207$, $p^{(B)}_3\approx 0.9716$, $p^{(C)}_1
\approx 0.2379$, $p^{(C)}_2 \approx 0.2031$, $p^{(C)}_3\approx 0.2044$,
$p^{(D)}_1 \approx 0.3760$, and $p^{(D)}_2 \approx 0.4200$. The discussed
values are obtained using numerical calculations. It can be seen that
these values for the probabilities imply that the pairs of qubits
$\hat{\rho}^{(A)}$, $\hat{\rho}^{(B)}$ and $\hat{\rho}^{(C)}$,
$\hat{\rho}^{(D)}$ have the same purity, and that it is close to the unity.
States reaching this upper bound are shown in the Appendix~\ref{A}. The~lower bound
corresponds to $p^{(B), (C), (D)}_j=1/2$, with $j=1,2,3$.

We note that different parameterizations do not lead to the same maxima of the areas. For
example, the $A,B,D$ parameterization allows for greater purity of qubits, thus
yielding a lower total sum of areas. If one requires the purity of the qutrit to be equal to $1$, and
equal purities for $\hat{\rho}^{(A)}$ and $\hat{\rho}^{(B)}$, and for
$\hat{\rho}^{(C)}$ and $\hat{\rho}^{(D)}$, this yields a maximum value of
$S\approx 8.095$.

\section{Separability Properties of the Two-Qubit Composite Systems}
\label{S3}
Given the density matrix of two qubits in the form $\rho_{m_1, m_2,
m_1^\prime, m_2^\prime}$, ($m_1,m_2,m_1^\prime, m_2^\prime=\pm 1/2$), i.e.,
\begin{equation}
\hat{\rho}(1,2)= \left( \begin{array}{cccc}
\rho_{\frac{1}{2},\frac{1}{2},\frac{1}{2},\frac{1}{2}} & \rho_{\frac{1}{2},
\frac{1}{2},\frac{1}{2},-\frac{1}{2}} & \rho_{\frac{1}{2},\frac{1}{2},
-\frac{1}{2},\frac{1}{2}} & \rho_{\frac{1}{2},\frac{1}{2},-\frac{1}{2},-\frac{1}{2}} \\
\rho_{\frac{1}{2},-\frac{1}{2},\frac{1}{2},\frac{1}{2}}
& \rho_{\frac{1}{2},-\frac{1}{2},\frac{1}{2},-\frac{1}{2}}
& \rho_{\frac{1}{2},-\frac{1}{2},-\frac{1}{2},\frac{1}{2}}
& \rho_{\frac{1}{2},-\frac{1}{2},-\frac{1}{2},-\frac{1}{2}} \\
\rho_{-\frac{1}{2},\frac{1}{2},\frac{1}{2},\frac{1}{2}}
& \rho_{-\frac{1}{2},\frac{1}{2},\frac{1}{2},-\frac{1}{2}}
& \rho_{-\frac{1}{2},\frac{1}{2},-\frac{1}{2},\frac{1}{2}}
& \rho_{-\frac{1}{2},\frac{1}{2},-\frac{1}{2},-\frac{1}{2}} \\
\rho_{-\frac{1}{2},-\frac{1}{2},\frac{1}{2},\frac{1}{2}}
& \rho_{-\frac{1}{2},-\frac{1}{2},\frac{1}{2},-\frac{1}{2}}
& \rho_{-\frac{1}{2},-\frac{1}{2},-\frac{1}{2},\frac{1}{2}}
 & \rho_{-\frac{1}{2},-\frac{1}{2},-\frac{1}{2},-\frac{1}{2}}
\end{array}\right) \, ,
\label{twoq}
\end{equation}
we consider two different situations for the two-qubit systems. The first one
where two states are not available or forbidden, while in the second case only
one is inaccessible. We analyze these different possibilities below.

\subsection{Two Inaccessible States}

A two-qubit density matrix~(Equation \ref{twoq}), in which two of the states (with
$m_1,m_2=1/2,1/2$ and $-1/2,-1/2$) are inaccessible, can be expressed as
\begin{equation}
\hat{\rho}(1,2)=\left( \begin{array}{cccc}
0 & 0 & 0 & 0 \\
0 & \rho_{11} & \rho_{12} & 0 \\
0 & \rho_{21} & \rho_{22} & 0 \\
0 & 0 & 0 & 0
\end{array} \right) \, .
\label{eq21}
\end{equation}

This state can be related to an equilibrium density operator
$\hat{\rho}=e^{-\hat{H}/T}/{\rm Tr}(e^{-\hat{H}/T})$, where the Hamiltonian
has very large first and fourth eigenvalues in comparison with the other two,
so that the transitions to the corresponding eigenstates are forbidden. Since
the qubit density matrix is expressed in terms of the probabilities $p_1$,
$p_2$, and $p_3$, Equation~(\ref{eq21}) can be written as
\begin{equation}
\hat{\rho}(1,2)=\left( \begin{array}{cccc}
0 & 0 & 0 & 0 \\
0 & p_3 & p_1-1/2-i (p_2-1/2) & 0 \\
0 & p_1-1/2+i (p_2-1/2) & 1-p_3 & 0 \\
0 & 0 & 0 & 0
\end{array} \right) \, .
\label{example1}
\end{equation}

{Next, we present a quantification of the entanglement by means of the
negativity~\cite{kowski} and concurrence~\cite{wooters,wooters1} concepts. The
negativity is defined by the sum of the absolute values of the negative
eigenvalues of the ppt density matrix $\hat{\rho}^{PT}$, that is, $\mathcal{N} (\hat{\rho})=
\sum_k\vert \lambda_{k}{}^{(-)}\vert$. Thus, one constructs the partial
transpose density matrix in the probability representation, which has
eigenvalues $\lambda_1=p_3$, $\lambda_2=1-p_3$,
$\lambda_3=\sqrt{(p_1-1/2)^2+(p_2-1/2)^2}$, $\lambda_4=-\lambda_3$. These
probabilities satisfy Equation~(\ref{cons}), hence the negativity of the system is}
 \begin{equation}
 \mathcal{N}(\hat{\rho})=\sqrt{(p_1-1/2)^2+(p_2-1/2)^2} \, ,
 \label{nega1}
 \end{equation}
and we immediately see that, for special values of $p_1=p_2=1/2$, the
two-qubit state is separable. For~all the other values of the probabilities,
the state is entangled.

\begin{figure}[!]
\centering \subfigure[] {\includegraphics[scale=0.25]{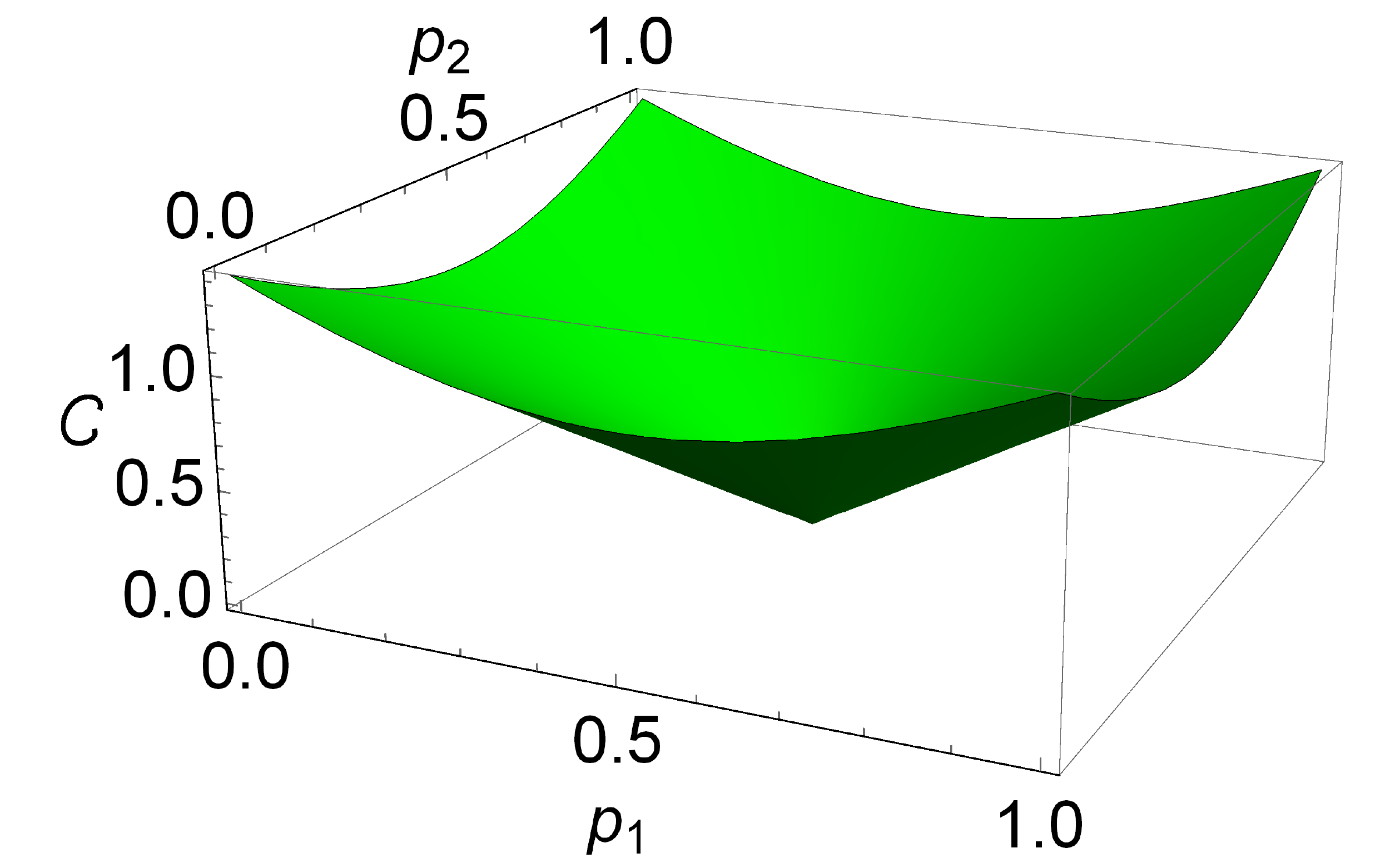}}
 \subfigure[] {\includegraphics[scale=0.25]{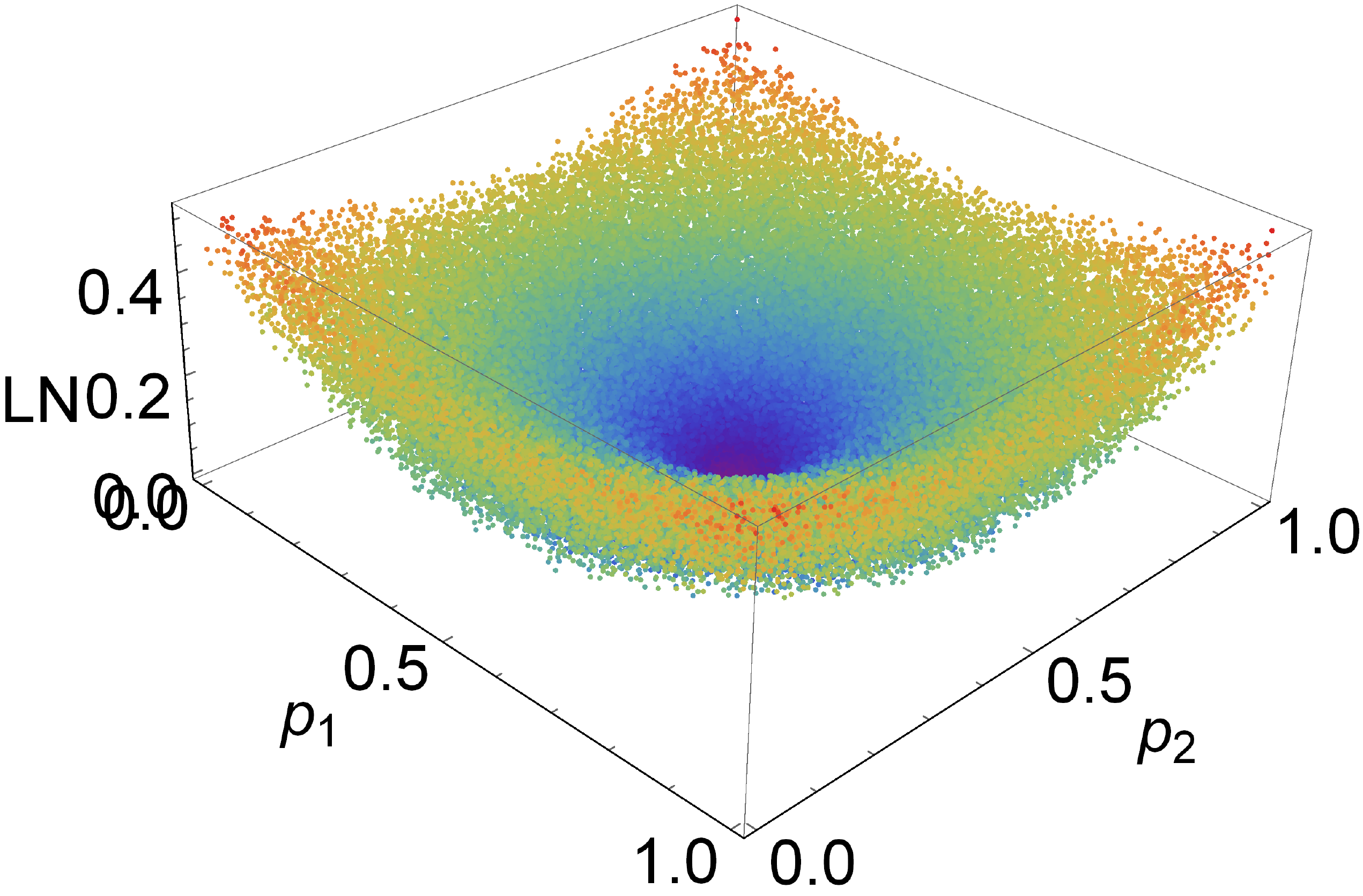}} \vspace{-8pt}
 \caption{(\textbf{a}) Quantum concurrence for $\hat{\rho}(1,2)$;
and (\textbf{b}) the numeric logarithmic negativity for the density matrix
$\hat{\rho}_{1}(1,2)$ in terms of the corresponding probabilities $p_1$ and
$p_2$.}
\label{fig:entang}
\end{figure}

We obtain the concurrence of the system by calculating the square root of the
eigenvalues of the matrix $\hat{\rho}\hat{\rho}^\prime$, where
$\hat{\rho}^\prime=(\hat{\sigma}_y\otimes
\hat{\sigma}_y)\hat{\rho}^*(\hat{\sigma}_y\otimes \hat{\sigma}_y)$, $\hat{\rho}^*$ is the complex conjugate of $\hat{\rho}$, and with $\hat{\sigma}_y$ being the Pauli matrix. The square root of
the eigenvalues of such a matrix in descending order ($\eta_1$, $\eta_2$,
$\eta_3$, and~$\eta_4$) defines the concurrence
$\mathcal{C}=\max\left(0,\eta_1-\eta_2-\eta_3-\eta_4\right)$. Given the state
of Equation~(\ref{example1}), these~are
\[
\eta_{1,2}=\sqrt{p_3(1-p_3)}\pm\sqrt{(p_1-1/2)^2+(p_2-1/2)^2}\, , \quad
\eta_{3,4}=0 \, ;
\]
thus, the concurrence of the state is
\begin{equation}
\mathcal{C}=2 \sqrt{(p_1-1/2)^2+(p_2-1/2)^2}=2 \vert \hat{\rho}(1,2)_{23}\vert
= 2 \mathcal{N}(\hat{\rho}) \, ;\label{conc1}
\end{equation}
it is shown in Figure~\ref{fig:entang}a. Here, we see that the concurrence is zero when $p_1=p_2=1/2$, i.e., when the state is diagonal, and has a maximum value when both probabilities are equal to one of the extreme values, zero or one; this corresponds to the different states, where $\hat{\rho}(1,2)_{2,3}$ is either $(-1\pm i)/2$ or $(1\pm i)/2$, and the inequality $p_3(1-p_3)\leq 3/4$ is satisfied.

We can also analyze the separability of the states in terms of the
square areas. This can be done by taking such four matrix elements that are
different from zero as a qubit. In the case where the system is separable,
$\mathcal{N}(\hat{\rho})=0$; $p_1=p_2=1/2$, one has from Equation~(\ref{cons}) that
the value of the other probability is unrestricted $0\leq p_3 \leq 1$.
However, the sum of the square areas $S=p_3 (4 p_3-5)+3$ can
take values $3/2\leq S \leq 5/2$, while if the system is entangled the
probabilities $(p_1-1/2)^2+(p_2-1/2)^2=\mathcal{N}^2(\hat{\rho})$ are located
within a circle of radius equal to the negativity of the system, and we should
have $1/2 (1- \sqrt{1 - 4\, \mathcal{N}^2(\hat{\rho})}) \leq p_3 \leq 1/2 (1 +
\sqrt{1 - 4\, \mathcal{N}^2(\hat{\rho})})$. Since the negativity takes a value
between 0 and $1/2$, we have $0\leq p_3 \leq 1/2$. From these arguments, one
can see that the sum of the square areas can take any value between
3/2 and 3. The interval $(5/2,\, 3]$ for $S$ provides the possibility for
experimental checkup of the entanglement of the system $\hat{\rho}(1,2)$ in
terms of probabilities.

Now, consider the case where the state is given by the density matrix
\begin{equation}
\hat{\rho}(1,2)=\left( \begin{array}{cccc}
p_3 & 0 & 0 & p_1-1/2-i (p_2-1/2) \\
0 & 0 & 0 & 0 \\
0 & 0 & 0 & 0 \\
p_1-1/2+i (p_2-1/2) & 0 & 0 & 1-p_3
\end{array} \right) \, .
\label{example2}
\end{equation}

As in the previous case, the state can be written in the form
$\hat{\rho}=e^{-\hat{H}/T}/{\rm Tr}(e^{-\hat{H}/T})$, where the Hamiltonian
has very large second and third eigenvalues compared with the other two.

The eigenvalues of the partial transpose are the same as in the previous
example, so the negativity is also given by Equation~(\ref{nega1}), and the
concurrence provides the same result of Equation~(\ref{conc1}). Hence, one can
conclude that there is entanglement for $p_{1,2}\neq 1/2$.

\subsection{One Inaccessible State}

In this case, the density operator can be described by a 3 $\times$ 3-matrix
inside the general \mbox{4 $\times$ 4-matrix}.  To establish its qubit representation,
we consider, following~\cite{chernega17c}, density matrices of the form
\begin{equation}
\hat{\rho}_{1} (1,2) =  \left( \begin{array}{cccc}
R_{11} & R_{12} & R_{13} & 0 \\
R_{21} & R_{22} & R_{23} & 0\\
R_{31} & R_{32} & R_{33} & 0 \\
0 & 0 & 0 & 0
\end{array}\right)  \, , \qquad
\hat{\rho}_{2} (1,2)=  \left( \begin{array}{cccc}
0 & 0 & 0 & 0 \\
0 & R_{11} & R_{12} & R_{13} \\
0 & R_{21} & R_{22} & R_{23} \\
0 & R_{31} & R_{32} & R_{33}
\end{array}\right)\, ,
\label{3by3}
\end{equation}
where the matrix $\hat{R}$ with elements $R_{jk}$; $j,k=1,2,3$ is the qutrit
density matrix. Since the qutrit is given in the probability representation by
~Equation (\ref{qutrit}), the two-qubit system represented by
$\hat{\rho}{(1,2)}$ can be also expressed in terms of probabilities.

To study the properties of entanglement, we use the Peres--Horodecki criterion
and construct the positive partial transpose matrix $\hat{\rho}^{PT} (1,2)$
with the map $T_2= I\otimes T$, where $T$ stands for the transpose operator,
which yields to two matrices; one for each matrix in Equation~(\ref{3by3}),
\begin{equation}
\hat{\rho}^{PT}_{1} (1,2) =  \left( \begin{array}{cccc}
R_{11} & R_{21} & R_{13} & R_{23} \\
R_{12} & R_{22} & 0 & 0\\
R_{31} & 0 & R_{33} & 0 \\
R_{32} & 0 & 0 & 0
\end{array}\right) \, , \qquad
\hat{\rho}^{PT}_{2} (1,2)=  \left( \begin{array}{cccc}
0 & 0 & 0 & R_{12} \\
0 & R_{11} & 0 & R_{13} \\
0 & 0 & R_{22} & R_{32} \\
R_{21} & R_{31} & R_{23} & R_{33}
\end{array}\right) \, .
\label{ppt}
\end{equation}

The criterion reads: If any of the eigenvalues of the matrices~in Equation (\ref{ppt}) is
negative, then the states described by the matrices in Equation~(\ref{3by3}) are
entangled.

As an example, we consider the state $\hat{\rho}_{1}(1,2)$ of
Equation~(\ref{3by3}), with each one of its elements described by the probabilities
as in Equation~(\ref{qutrit}). This time, the square root of the eigenvalues of
$\hat{\rho}\hat{\rho}^\prime$ are
$\eta_{1,2}=\sqrt{(1-p_3^{(A)})(1-p_3^{(B)})}\pm \vert D \vert$ and
$\eta_{3,4}=0$. From this, the concurrence is
\begin{equation}
\mathcal{C}=2 \vert D \vert=2 \sqrt{\left(p_1^{(D)}-1/2\right)^2+\left(p_2^{(D)}-1/2\right)^2} \, ,
\end{equation}
implying entanglement when $D\neq 0$ $\left(p_{1,2}^{(D)}\neq 1/2\right)$.

In addition, the separability condition $\mathcal{C}=0$ implies that the sum of
the square areas for qubit $\hat{\rho}^{(D)}$ is restricted to values
between $3/2$ and $5/2$. Thus, in the separable case, the value of the sum of
the areas of the triads is bounded by the range $9/2\leq S
\leq 8$. The value of the sum $S=8$ is attained when
$p_1^{(B)}=p_2^{(B)}=p_1^{(D)}=p_2^{(D)}=1/2$, $p_3^{(B)}=1$ and
$p_1^{(C)}=p_2^{(C)}=p_3^{(C)}=(3+\sqrt{3})/6$.

In the case of $\hat{\rho}_{2}$, the square root of the eigenvalues of
$\hat{\rho}\hat{\rho}^\prime$ are $\eta_{1,2}=\sqrt{(p_3^{(A)}+p_3^{(B)}-1)
(1-p_3^{(B)})}\pm \vert B \vert$ and $\eta_{3,4}=0$. From these values, the
concurrence is calculated to~be
\begin{equation}
\mathcal{C}=2 \vert B \vert= 2 \sqrt{\left(p_1^{(B)}-1/2\right)^2
+\left(p_2^{(B)}-1/2\right)^2} \, ,
\end{equation}
which means that the system is separable when $B=0$. In addition, we can notice that
the sum of the square areas has the same bounds as in the previous case
($9/2\leq S \leq 8$).

Finally, we consider the state
\begin{equation}
\hat{\rho}=\left(\begin{array}{cccc}
R_{11} & 0 & R_{12} & R_{13} \\
0 & 0 & 0 & 0 \\
R_{21} & 0 & R_{22} & R_{23} \\
R_{31} & 0 & R_{32} & R_{33}
\end{array}\right) \, .
\end{equation}

We found the eigenvalues
$\eta_{1,2}=\sqrt{(p_3^{(A)}+p_3^{(B)}-1)(1-p_3^{(A)})}\pm \vert A \vert$,
$\eta_{3,4}=0$, and the concurrence of the form
\begin{equation}
\mathcal{C}=2 \vert A \vert = 2 \sqrt{\left(p_1^{(A)}-1/2\right)^2+\left(p_2^{(A)}-1/2\right)^2}\, .
\end{equation}

It can be shown that the sum of the square areas for the separable case
has the bounds $9/2\leq S \leq (57+\sqrt{17})/8$, where the maximum is
obtained when $\hat{\rho}$ describes the pure state, the qubits have the same
purity ${\rm Tr}(\hat{\rho}^{(A)2})={\rm Tr}(\hat{\rho}^{(B)2})$ and ${\rm
Tr}(\hat{\rho}^{(C)2})={\rm Tr}(\hat{\rho}^{(D)2})$, with one of these
purities equal to 1. This can be attained for the probabilities
$p_1^{(B)}=p_2^{(B)}=p_1^{(C)}=p_2^{(C)}=1/2$, $p_3^{(B)}=1/2 (1 \pm \sqrt{1/2
+ 3/(2 \sqrt{17})})$, $p_3^{(C)}=0$, and $p_{1,2}^{(D)}=1/2 \mp 1/4 \sqrt{1 -
3/\sqrt{17}}$.

The entanglement properties of the physical system described by the density
matrix, in which the third row and third column vanish, are analogous to those
of $\hat{\rho}_1(1,2)$. In this case, all the expressions for the concurrence
have the same analytic form as for the two inaccessible states; they are also
depicted in Figure~\ref{fig:entang}a.  The separability of the systems, when
$D=0$ or $B=0$ or $A=0$, can be checked using the partial transpose procedure.
In all these cases, the eigenvalues of the partial transpose are equal to the
nonnegative eigenvalues of the original density matrix, so the negativity
vanishes.

When the system state is not separable, the calculation of the negativity can only
be done numerically. In Figure~\ref{fig:entang}b, we illustrate the behavior of
the logarithmic negativity $LN(\hat{\rho})= \ln{(2 \,
\mathcal{N}(\hat{\rho})+1)}$ for the system $\hat{\rho}_1(1,2)$. We notice that the logarithmic negativity is zero for the values $p_1=p_2=1/2$, and the state is diagonal. In addition, the logarithmic negativity has a maximum value when both probabilities correspond to an extreme value, zero or one. The probabilities $p_1$ and $p_2$ at the extremal values of the logarithmic negativity are the same as the ones for the concurrence.

\section{Example}
\label{S4}
We now consider the coherent state for spin $J=1$ (cf., e.g.,~\cite{radcliffe})
\[
\vert \zeta \rangle = \frac{1}{1+\vert \zeta \vert^2}(\vert 1,-1 \rangle+ \sqrt{2}\, \zeta \vert 1,0 \rangle+ \zeta^2 \vert 1,1\rangle) \, ,
\]
where $\zeta$ is a complex parameter given in terms of the polar and azimuthal angles of the Bloch sphere. This state is interesting in regards to the Einstein-Podolsky-Rosen paradox.
experiment when taken as a symmetric state of two spin-$1/2$ particles. The fact that one may determine the separability or entanglement of the two particles by only measuring $2$ components of $\hat{J}$ in a series of runs may provide an advantage to experimental setups.

This pure state defines the following qubit probabilities in terms of the mean values of the spin operators $\hat{J}_x$, $\hat{J}_y$, and $\hat{J}_z$
\begin{eqnarray}
\hspace{-0.3in}&&p_1^{(A)}=\frac{1}{4}(2+\langle \hat{J}_x \rangle^2-\langle \hat{J}_y \rangle^2),\hspace{0.4in} p_2^{(A)}=\frac{1}{2}(1+\langle \hat{J}_x \rangle \langle \hat{J}_y \rangle ), \hspace{0.65in} p_3^{(A)}=\frac{1}{4}(3-\langle \hat{J}_z \rangle)(1+\langle \hat{J}_z \rangle)\, , \nonumber \\
\hspace{-0.3in}&&p_1^{(B)}=\frac{1}{4} \left(2+\sqrt{2} \langle \hat{J}_x \rangle (1+\langle \hat{J}_z \rangle)\right),\ \  p_2^{(B)}=\frac{1}{4} \left(2+\sqrt{2} \langle \hat{J}_y \rangle (1+\langle \hat{J}_z \rangle)\right), \ \  p_3^{(B)}=\frac{1}{2}(1+ \langle \hat{J}_z\rangle^2)\, ,  \\ \label{probs}
\hspace{-0.3in}&&p_1^{(D)}=\frac{1}{4} \left(2+\sqrt{2} \langle \hat{J}_x \rangle (1-\langle \hat{J}_z \rangle)\right),\ \  p_2^{(D)}=\frac{1}{4} \left(2+\sqrt{2} \langle \hat{J}_y \rangle (1-\langle \hat{J}_z \rangle)\right), \ \  p_3^{(D)}=\frac{1}{2}(1-\langle \hat{J}_z \rangle)(1+\langle \hat{J}_z \rangle) \, ,  \nonumber
\end{eqnarray}
via which the classical probabilities can be measured experimentally. Although these expressions depend of the three mean values, the dependence can be reduced to only two by the property $\langle \hat{J}_x \rangle^2+\langle \hat{J}_y \rangle^2+\langle \hat{J}_z \rangle^2=1$. Given this, one can immediately check the constrictions for every one of the qubits~(Equation \ref{cons}), resulting in $0\leq \frac{1}{8} \left(\langle \hat{J}_z \rangle^4\mp2 \langle \hat{J}_z \rangle^3\pm2 \langle \hat{J}_z \rangle+1\right)\leq 1/4$, which can be reduced to the standard condition $-1\leq \langle \hat{J}_z \rangle \leq 1$. Furthermore, the inequalities over the squares areas~(Equation~\ref{mal_qubits}) for every one of the three qubits $A$, $B$, and $D$, lead to the expressions
\begin{eqnarray}
&&\frac{3}{2}\leq \frac{5 \langle \hat{J}_z \rangle^4}{8}-\frac{5 \langle \hat{J}_z \rangle^3}{4}+\frac{1}{4} \langle \hat{J}_z \rangle^2 \left(-2 \langle \hat{J}_x \rangle \langle \hat{J}_y \rangle+\langle \hat{J}_y \rangle^2-1\right)+\frac{1}{4} \langle \hat{J}_z \rangle (2 \langle \hat{J}_y \rangle (\langle \hat{J}_x \rangle-\langle \hat{J}_y \rangle)+5)+ \nonumber \\
&&\hspace{1.5in}
\frac{1}{8} \left(17-2 \langle \hat{J}_y \rangle \left(2 \langle \hat{J}_x \rangle \left(\langle \hat{J}_y \rangle^2-1\right)+\langle \hat{J}_y \rangle\right)\right) \leq 3 \, , \nonumber \\
&&\frac{3}{2} \leq \frac{\langle \hat{J}_z \rangle^4}{2}+\frac{1}{4} \langle \hat{J}_z \rangle^3 \left(\sqrt{2} \langle \hat{J}_x \rangle+\sqrt{2} \langle \hat{J}_y \rangle \mp 4\right)+\frac{1}{4} \langle \hat{J}_z \rangle^2 \left(\langle \hat{J}_x \rangle \left(\langle \hat{J}_y \rangle \pm\sqrt{2}\right) \pm \sqrt{2} \langle \hat{J}_y \rangle\right)\pm\nonumber \\ 
&&\hspace{1.5in}\langle \hat{J}_z \rangle \left(\frac{\langle \hat{J}_x \rangle \langle \hat{J}_y \rangle}{2}+1\right)+\frac{\langle \hat{J}_x \rangle \langle \hat{J}_y \rangle}{4}+2 \leq 3 \, .
\label{sum_q}
\end{eqnarray}

Again, these inequalities are constrained to $\langle \hat{J}_x \rangle =\pm \sqrt{1-\langle \hat{J}_y \rangle^2-\langle \hat{J}_z \rangle^2}$. On the other hand, the sum of the square areas~(Equation \ref{sums}) define the following inequality:
\begin{eqnarray}
\frac{9}{2} \leq \frac{13 \langle \hat{J}_z \rangle^4}{8}+\langle \hat{J}_z \rangle^3 \left(\frac{\langle \hat{J}_x \rangle}{\sqrt{2}}+\frac{\langle \hat{J}_y \rangle}{\sqrt{2}}-\frac{5}{4}\right)+\frac{1}{4} \left(\langle \hat{J}_y \rangle^2-1\right) \langle \hat{J}_z \rangle^2+\frac{1}{4} \langle \hat{J}_z \rangle (2 \langle \hat{J}_y \rangle (\langle \hat{J}_x \rangle-\langle \hat{J}_y \rangle)+5)+ \nonumber \\
\frac{1}{8} \left(49-2 \langle \hat{J}_y \rangle \left(2 \langle \hat{J}_x \rangle \left(\langle \hat{J}_y \rangle^2-2\right)+\langle \hat{J}_y \rangle\right)\right) \lesssim 8.095 \, .
\label{sum_t}
\end{eqnarray}

\begin{figure}[!]
\centering 
\subfigure[] {\includegraphics[scale=0.22]{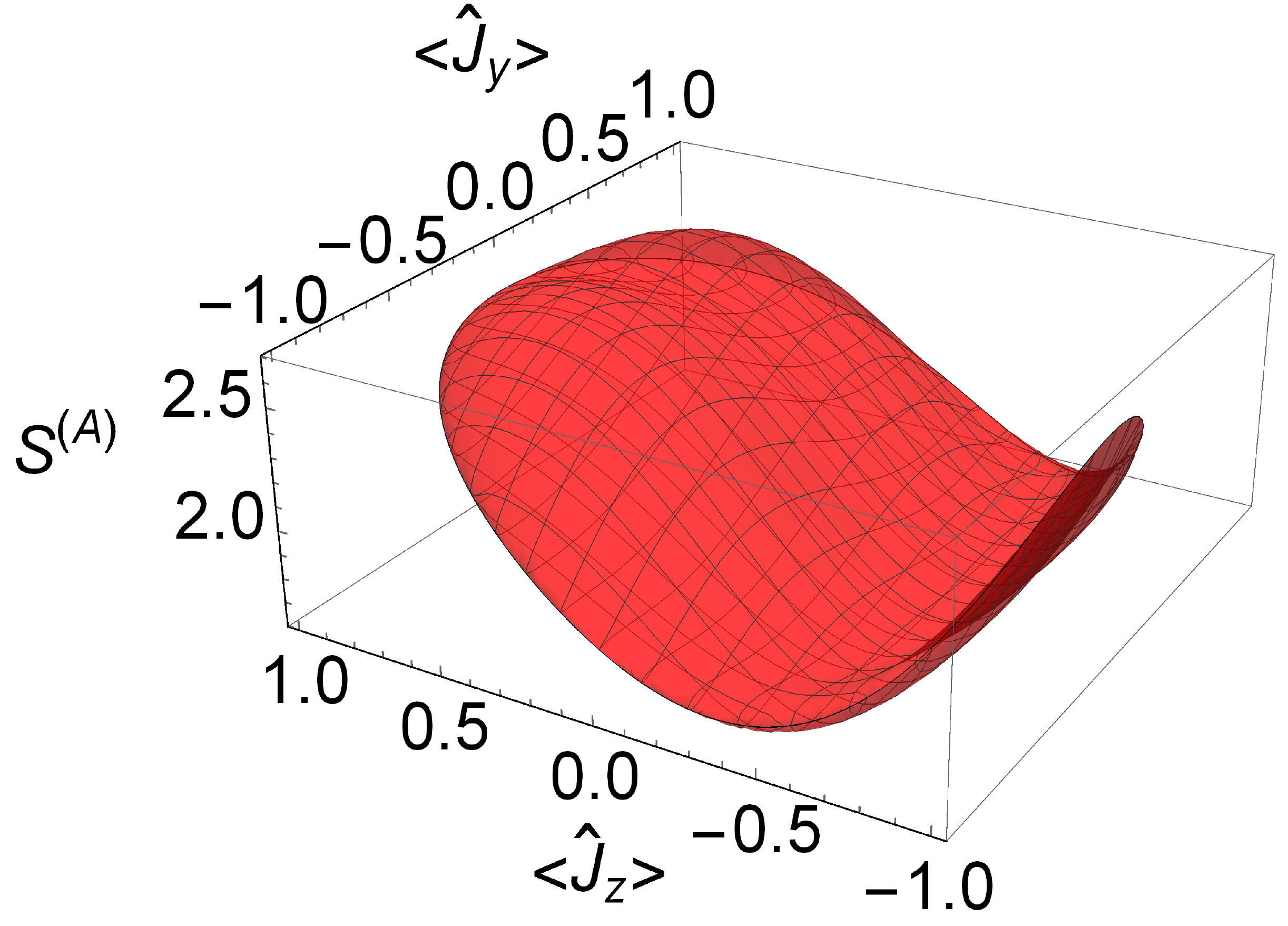}\label{area_a}} 
\ \subfigure[] {\includegraphics[scale=0.22]{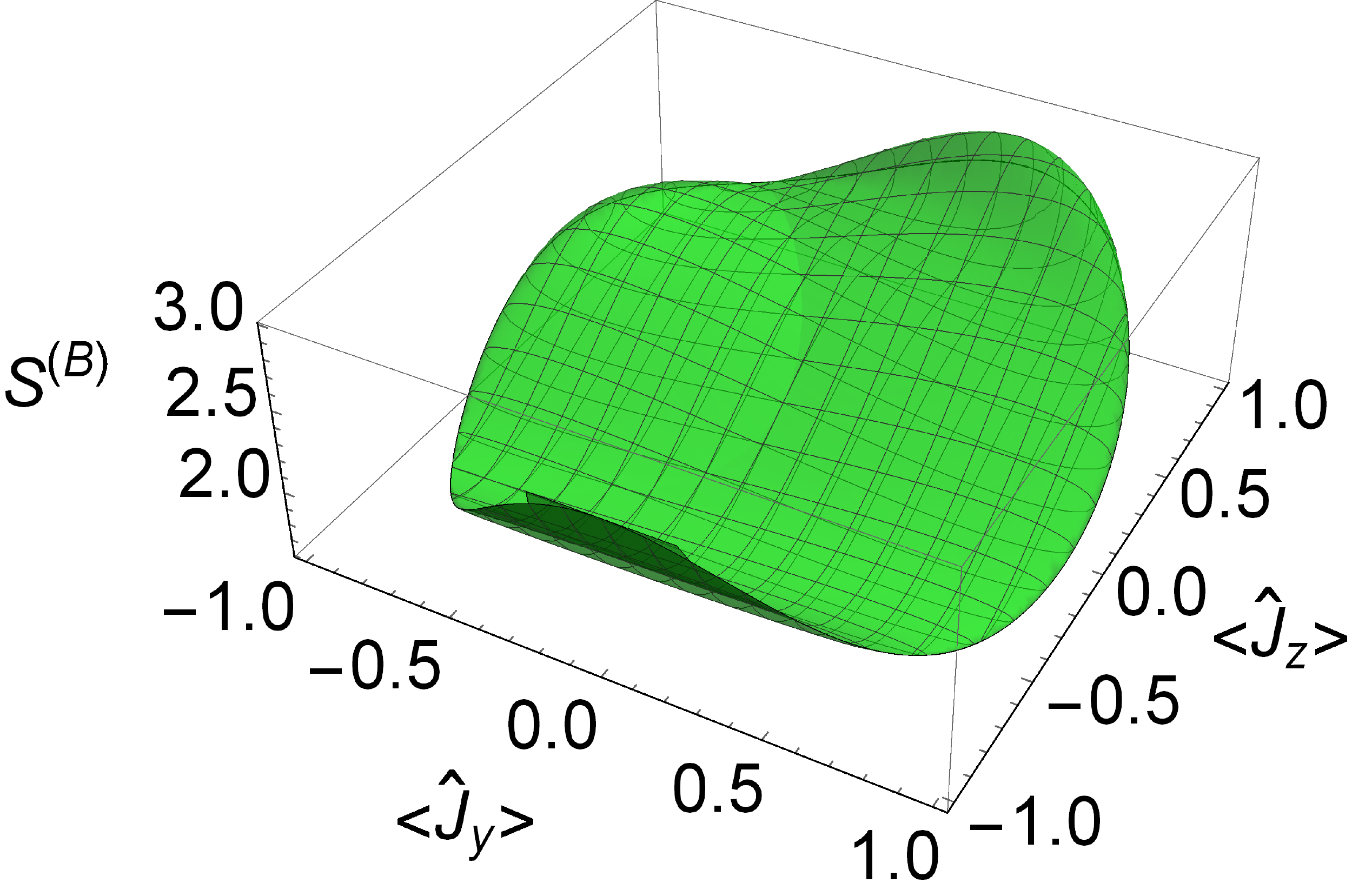}\label{area_b}} 
\ \subfigure[] {\includegraphics[scale=0.22]{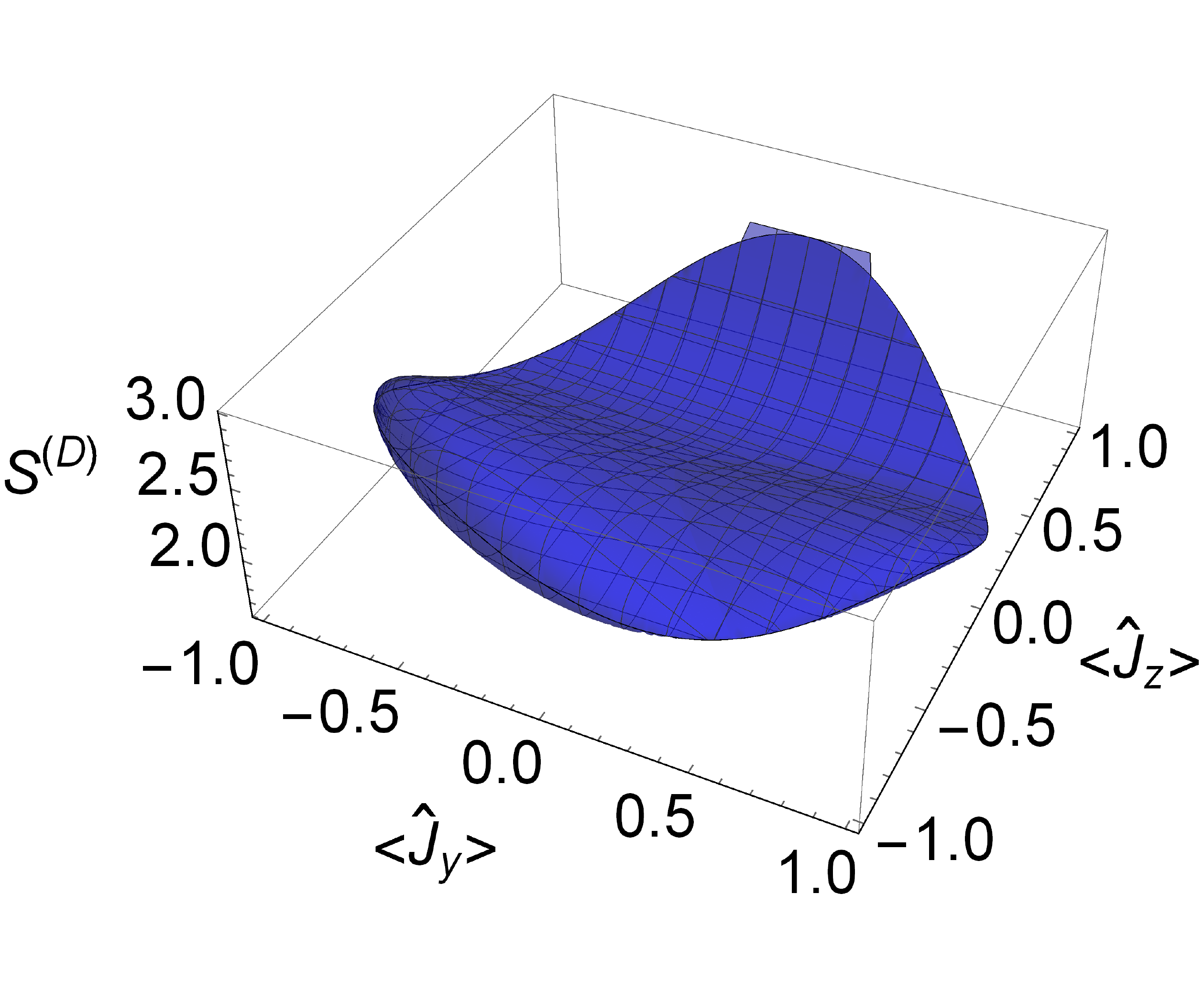}\label{area_d}} 
\ \subfigure[] {\includegraphics[scale=0.22]{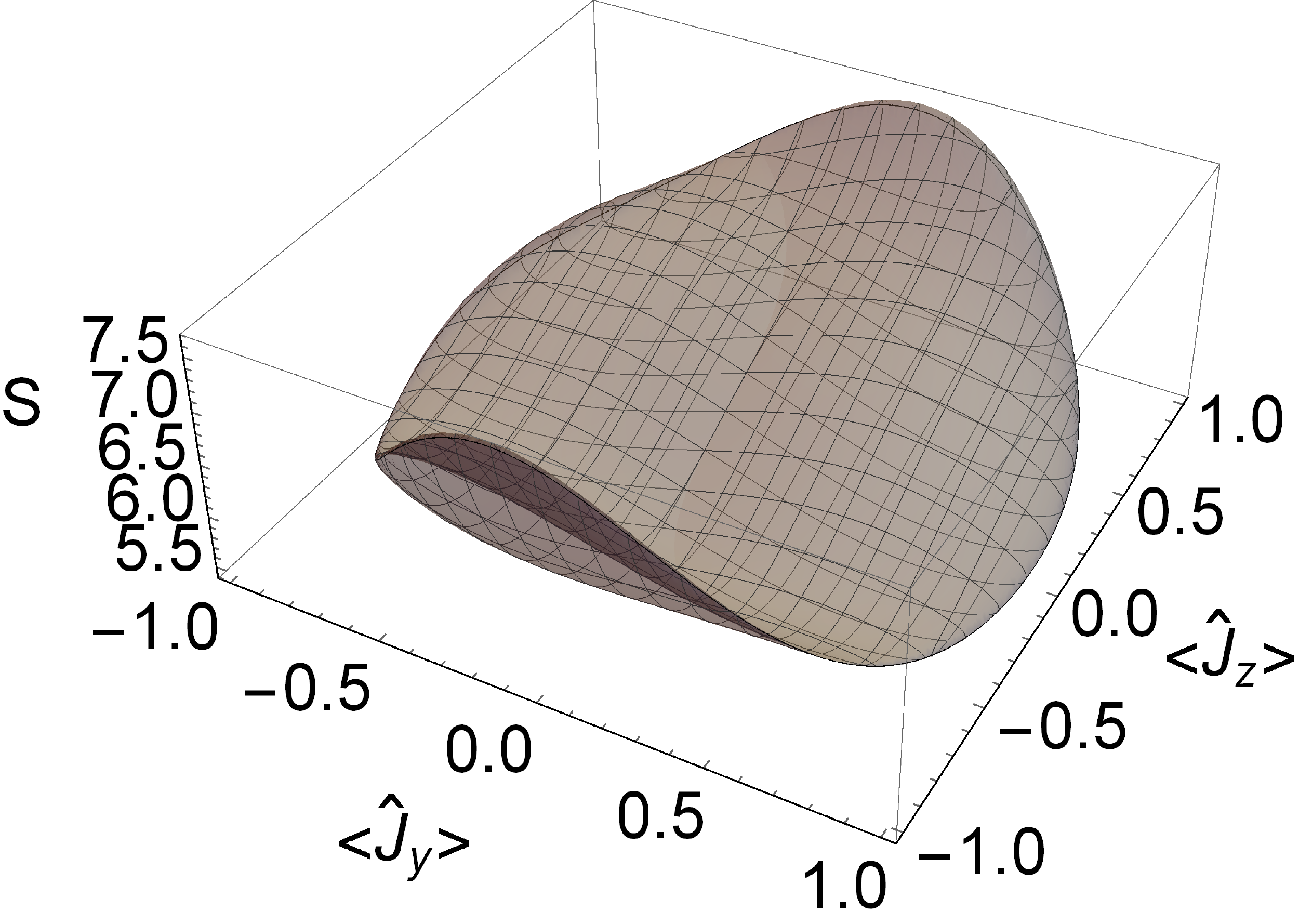}\label{area_t}} 
\caption{Sum of the square areas for the qubits: (\textbf{a})\, $\hat{\rho}^{(A)}$; (\textbf{b})\, $\hat{\rho}^{(B)}$; and (\textbf{c})\, $\hat{\rho}^{(D)}$. (\textbf{d}) Total sum of the areas $S$. All these functions depend of the mean values of the spin operators $\hat{J}_y$ and $\hat{J}_z$ of the coherent state $\vert \zeta \rangle$.}
\label{area}
\end{figure}

As the coherent state is very particular, the inequalities discussed above can be further reduced. In Figure~\ref{area_a}--\ref{area_d},   the allowed values for the sum of the square areas for the qubits $\hat{\rho}^{(A)}$, $\hat{\rho}^{(B)}$, and $\hat{\rho}^{(D)}$, defined by the coherent state $\vert \zeta \rangle$, are plotted in terms of the mean values  $\langle \hat{J}_y \rangle$ and $\langle \hat{J}_z \rangle$. As  can be seen, the possible values for these areas satisfy the condition~in~Equation (\ref{mal_qubits}). In Figure~\ref{area}d, the sum $S$ of the areas is also evaluated and the limits $(9/2,8.095)$ can be checked.

Finally, one can conclude that the conditions in~Equations ($\ref{sum_q}$) and (\ref{sum_t}) can be used as a control to check the experimental measurement of the mean values of the observables $\hat{J}_x$, $\hat{J}_y$, and $\hat{J}_z$.

\section{Conclusions}
\label{S5}
In this paper, we have used classical probabilities to describe quantum states, an approach which may provide a better understanding of quantum entanglement: the fact that this purely quantum phenomenon may be described by classical measurable probabilities seems remarkable. That the separability or entanglement of two-qubit systems can be described in purely classical terms has also been shown recently \cite{mancini}, where the classical Fisher metric on phase space is shown to give the same (qualitative) results as the quantum Fisher metric.

The definition of the Malevich squares and their areas is presented as a new approach to describe geometrically the qudit quantum state. In particular, the different limits for the sum of the square areas are obtained for general qubit and qutrit systems. We show some of the inequalities associated with the different areas for a spin-1 coherent state as an example of the applicability of our approach. The~possible use of these expressions as a control for experimental data is also addressed.

By means of this probabilistic construction of quantum mechanics, we present
the study of the linear entropy of general qubit and qutrit systems. In both
cases, the entropy is written in terms of classical probability distributions,
and their geometrical interpretation is discussed. In the qutrit case, one can
see that the linear entropy of the system is determined by the sum of the
entropies of its component qubits.

In addition, we constructed in explicit form the density matrix of some separable and
entangled states of two qubits in terms of fair classical probability
distributions. We obtained the characteristics of the entanglement, such as
the concurrence and numeric logarithmic negativity, as functions of the
probability distributions. The paradigmatic examples of the entangled states
correspond to eigenstates of degenerate two-qubit Hamiltonians, which are
defined in terms of three probabilities for the qubit state or eight
probabilities for the qutrit state.  In the latter case, these are selected
from twelve dichotomous probability distributions. In a future work, we extend
the procedure given here to multipartite systems.

We presented the geometrical picture of the entanglement in terms of triads
of squares and found the areas of the squares for entangled states.
It is worth noting that, when there is one or two inaccessible states for
the two-qubit system, its entanglement properties are determined in terms of
one or two spin-1/2 probability distributions. We always found an interval for
$S$ that provides the possibility for an experimental checkup of the system
entanglement in terms of probabilities.

To conclude, we emphasize that, in the probability representation of quantum
states, completely quantum phenomena such as Bell correlations in two-qubit
systems can be described using only properties of classical probability
distributions associated with probability
interferences~\cite{KhrPr,KhrFPh,KhrQI,khrebook} and nonlinear superposition
rules~\cite{Chern1,MAPS,chern2} for the probabilities determining the qudit
states.

\vspace{6pt}

\section*{Acknowledments}
 This work was partially supported by CONACyT--Mexico under
Project No.~238494 and DGAPA--UNAM under Project No.~IN101217. The work of V.I.M. and J.A.L.-S. was partially performed at the Moscow Institute of Physics and Technology, where V.I.M. was supported by the Russian Science Foundation under Project No. 16-11-00084. In addition, V.I.M. acknowledges the partial support of the Tomsk State University Competitiveness Improvement Program. M.A.M. and V.I.M. acknowledge the hospitality provided by the Institute for Nuclear Sciences, UNAM, Mexico. 

\appendix
\section{Upper Bound for the Sum of the Square Areas}
\label{A}
Due to the requirement for qutrit to be the pure state, its density matrix
reads
\[
\hat{\rho}=\left(\begin{array}{ccc}
1-p_\beta^2-p_\gamma^2 & \sqrt{1-p_\beta^2-p_\gamma^2}\,  p_\beta \,
e^{-i \beta} & \sqrt{1-p_\beta^2-p_\gamma^2}\, p_\gamma \, e^{-i \gamma} \\
\sqrt{1-p_\beta^2-p_\gamma^2}\,  p_\beta\, e^{i \beta} & p_\beta^2
& p_\beta \, p_\gamma \, e^{i(\beta-\gamma)} \\
\sqrt{1-p_\beta^2-p_\gamma^2}\, p_\gamma \, e^{i \gamma}
& p_\beta\,  p_\gamma \, e^{-i(\beta-\gamma)} & p_\gamma^2
\end{array}\right)\, ,
\]
for a state in the spin $s=1$ representation,
\[
\vert \psi \rangle= \sqrt{1-p_\beta^2-p_\gamma^2} \vert 1 \rangle+p_\beta e^{i \beta}
\vert 0 \rangle+p_\gamma e^{i \gamma} \vert -1 \rangle \, .
\]

Maximizing the sum of the square areas $S$ with respect to $p_\beta$,
$p_\gamma$, $\beta$, and $\gamma$, we obtain the upper bound given in
Equation~(\ref{bb}) for the states determined by the parameters given by
$p_\beta\approx 0.1685$, $p_\gamma\approx 0.8759$, $\beta\approx 0.2749$, and
$\gamma\approx 3.9892$.


\end{document}